\documentclass[authoryear,1p,review,doublespacing,12pt]{elsarticle}
\usepackage{amssymb}
\usepackage{amsmath}
\usepackage{wasysym}
\usepackage{natbib}
\usepackage{graphicx}
\usepackage{xcolor}
\usepackage{subfigure}
\usepackage{siunitx}
\usepackage[pdftitle={Yongxiang Huang: AE},bookmarks=true,citecolor=red,colorlinks=false]{hyperref}

 \usepackage{ulem}

\renewcommand{\eqref}[1]{~(\ref{#1})}

\newcommand{\red}[1]{\textcolor{black}{#1}}

\journal{AE}

\begin{document}

\begin{frontmatter}


\title{Spatial statistics  of  atmospheric particulate matter in China}

\author[SHUa]{Shenghui Gao}
\author[SHUb]{Yangjun Wang}

\author[XMU]{Yongxiang Huang}
\ead{yongxianghuang@gmail.com}
\author[SHUa]{Quan Zhou}
\author[SHUa]{Zhiming Lu}
\author[SHUb]{Xiang Shi}
\author[SHUa]{Yulu Liu}

\cortext[corauthF]{Corresponding author. Email: yjwang326@shu.edu.cn.}

\address[SHUa]{Shanghai Institute of Applied Mathematics and Mechanics, Shanghai Key Laboratory of Mechanics in Energy Engineering,
Shanghai University, Shanghai 200072,  China}
\address[SHUb]{School of Environmental and Chemical Engineering, Shanghai University, Shanghai 200444, China}
\address[XMU]{State Key Laboratory of Marine Environmental Science,
Xiamen University, Xiamen 361102,  China}

\begin{abstract}
In this paper, the spatial dynamics of the atmospheric particulate matters (resp. PM$_{10}$ and PM$_{2.5}$) are studied using   turbulence methodologies. 
It is found experimentally that  the spatial correlation function $\rho(r)$ shows a log-law on the mesoscale range, i.e., $50\le r\le 500\,\si{km}$, with an experimental scaling exponent $\beta=0.45$. The  spatial structure function  shows a power-law behavior  on the mesoscale range $90\le r\le 500\,\si{km}$. The experimental scaling exponent $\zeta(q)$ is convex, showing that the intermittent correction is relevant in characterizing the spatial dynamic  of particulate matter. The measured singularity spectrum $f(\alpha)$ also shows its multifractal nature.  Experimentally, the particulate matter is more intermittent than the passive scalar, which could be partially due to the \red{mesoscale movements} of the atmosphere, and  also due to  local sources, such as local industry activities. 
 \end{abstract}

\begin{keyword} Particulate matter \sep Logarithm spatial correlation\sep Multifractality 
\end{keyword}

\end{frontmatter}

\pagebreak

\section{Introduction}

In recent decades, many cities in China have experienced heavy air pollution episodes leading to negative impacts on human health \citep{Streets2000AE,Chan2008AE,Matus2012GEC,Chen2013heavy,Wang2014ERL,Wang2014JGR,Zhang2014EI,Rohde2015P1}, and the air pollution have been one of the biggest problems in urban areas of many megacities in China. The Jing-Jin-Ji region, Yangtze River Delta region, Pearl River Delta region, Central China region and Cheng-Yu region, to list a few,  are the major polluted regions in China due to  highly densed population and high energy consumption. In order to improve the air quality, China government issued new national ambient air quality standards in 2012 and were to be implemented in Jan.2016. According to the new standards, air quality indices ranging from 0 to 50 and ranging from 51 to 100 represent excellent and good, respectively. However, air quality index equal to or above 101 means the air quality does not meet the national ambient air quality standards. Hourly observed concentration data for pollutants in numerous cities were released by the government. According to the new ambient air quality standards,  only 8 out of China's 74 biggest cities  met the government's air quality standards in 2014. Although not by much, the air quality in 2014 was better than in 2013: see
\href{http://www.mep.gov.cn/gkml/hbb/qt/201502/t20150202_295333.htm}{http://www.mep.gov.cn/}. {Particulate matter with an aerodynamic diameter $10\,\si{\mu m}$ or less, or
 PM$_{10}$ usually 
 dominantes  pollution episodes caused by dust storms}. Particulate matter with a diameter less than $2.5\,\mu$m,  or PM$_{2.5}$ usually could lead to more serious health problems for local residents than coarse particle due to easier inhalation. Furthermore, most haze episodes occurring in China are characterized by high concentrations of PM$_{2.5}$ in the ambient air. 

\cite{Rohde2015P1} applied the Kriging interpolation to four months of data to retrieve the pollution maps for eastern China and discovered that the greatest pollution occurs in the east. Air pollution episodes can cover a large region and are particularly intense in a northeast corridor that extends from outside of Shanghai to north of Beijing. Particulate matter is a very complicated mixture that comes from numerous emission sources. 
\red{
Industrial process was the dominant local contributor to PM2.5 concentration in the whole city of Shanghai except at the urban center where vehicle emissions contribute slightly more  \citep{Wang2014JGR}. Moreover, haze episode could be caused by the combination of anthropogenic emissions, unusual atmospheric circulation, the depression of strong cold air activities, and weak boundary layer ventilation \citep{Wang2014STE}. 
\cite{Wang2014ERL} concluded that the response of PM$_{2.5}$ to  meteorology possibly  changes  a feedback loop whereby planetary boundary layer dynamics amplify the initial perturbation of PM$_{2.5}$.} 

Note that the air pollution  occurred in the planetary boundary layer, where atmospheric turbulence is involved, 
showing a significant impact on the transport and dispersion of  pollution matter.  However, the effects of atmospheric turbulence are seldom studied in a multiscale view. A common behavior of the turbulence is the multiscaling, or multifractality, of the velocity field \citep{Frisch1995}.  In the view of the hydrodynamic turbulence,  a large range of spatial and temporal scales/freedoms are involved,  resulting in a cascade process in which the energy transfers from  large-scale structures to small-scale ones until the fluid viscosity converts the kinetic energy into heat.  This phenomenological Kolmogorov-Richardson energy cascade picture has been widely and successfully applied  in  multiple disciplinary fields, such as financial activity \citep{Schmitt1999,Ghashghaie1996a,Li2014PhysicaA}, crack of rock surfaces  \citep{Schmittbuhl1995JGR}, rainfall patterns \citep{Tessier1996}, etc. 

Specifically for  atmospheric turbulence,
due to the geometrical  constrain of the atmospheric movement, there exists several typical spatial scales. These include  the microscale (resp. $1\,\si{km}$ or less), showing  three-dimensional property   and synoptic scale (resp. up to $1000\,\si{km}$),    showing a two-dimensional feature \citep{Vallgren2011PRL}.  Between the microscale and synoptic scale, a large range of scale motion exists in  mesoscale structures (resp. from few dozens of km to few hundreds km).   The famous Kolmogorov $5/3$-law has been observed on the mesoscale range \citep{Nastrom1984Nature,Nastrom1985JAS} and agrees well with the above mentioned Kolmogorov-Richardson cascade prediction \citep{Vallgren2011PRL}.  Therefore, the air pollution indices, such as PM$_{2.5}$ could display a spatial scaling behavior since  they are advected mainly by these mesoscale structures. In this paper, we  employ the standard structure function analysis to retrieve the multiscale and multiscaling properties of the PM$_{10}$ and PM$_{2.5}$ to show the impact from the mesoscale atmospheric turbulence. 
 
\section{Data}\label{sec:data}
\begin{figure}[!htb]
\centering
\includegraphics[width=0.98\textwidth]{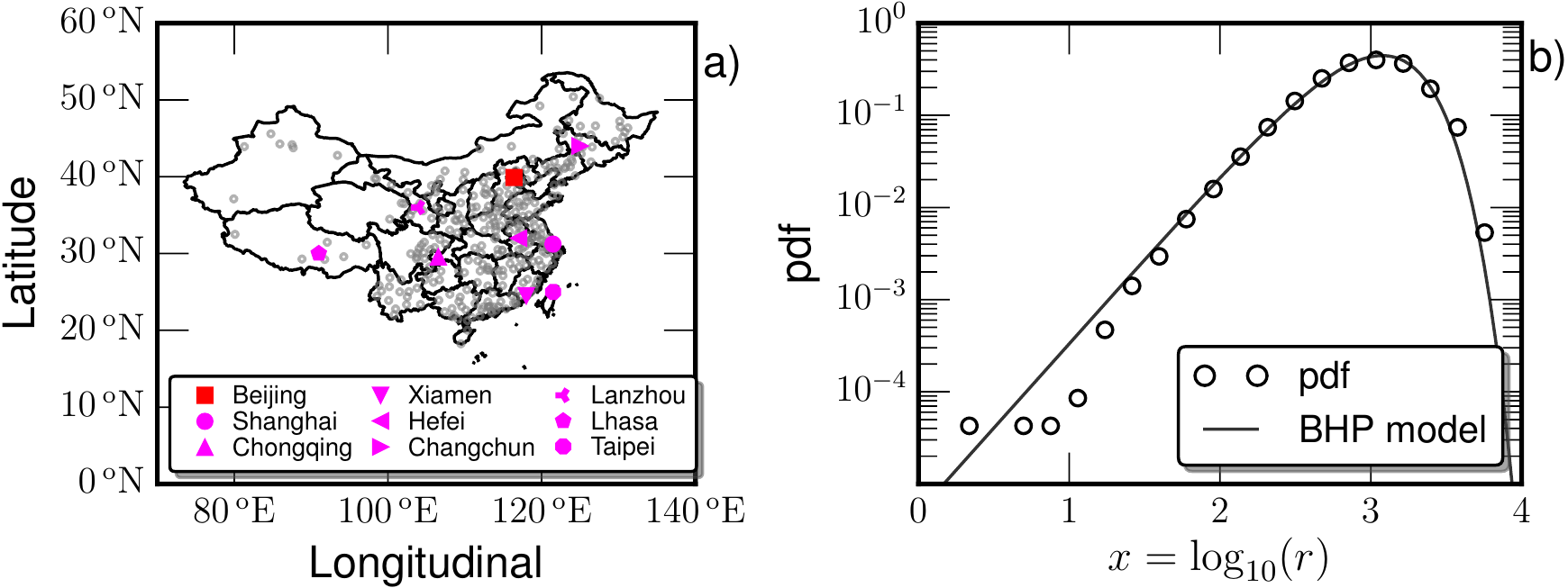}
\caption{a) The spatial distribution  of 305  monitor stations in different cities.  b) The distribution of the neighbour distances ($\ocircle$). The BHP model with parameters $b=0.938$, 
 and $K=2.14$ obtained numerically is illustrated by a solid line.}
\label{fig:cities} 
\end{figure}

\begin{figure}[!htb]
\centering
\includegraphics[width=0.95\textwidth]{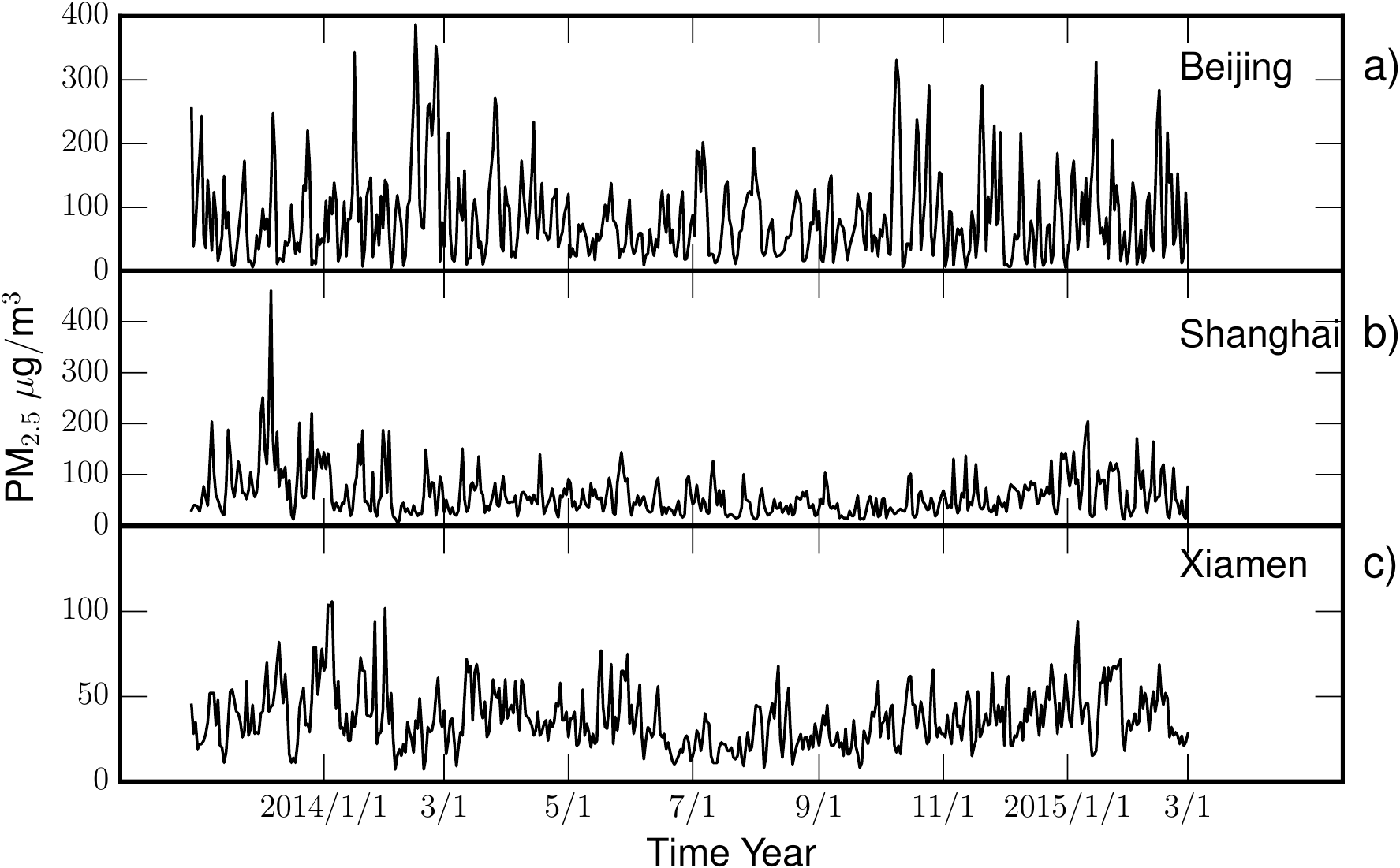}
\caption{Illustration of the recorded PM$_{2.5}$ in $\si{\mu g/m^3}$ of three typical cities: a) Beijing, b) Shanghai and c) Xiamen. Large variance is visible, showing the intermittency pattern. }
\label{fig:Example} 
\end{figure}

The hourly concentrations of PM2.5 and PM10  were released by the government (\href{http://www.cnemc.cn)}{http://www.cnemc.cn}). We processed these data into daily average concentrations to be used in this study.  
There are 305 monitor stations  belonging to different cities.  Fig.\ref{fig:cities}\,a) shows the spatial distribution of these monitor stations, which were monitored during the period  from 31 Dec. 2013 to 01 Mar. 2015, corresponding to 425 days for most of cities with several missing. In total, there are $82,755$ daily averaged data points. The neighbor distance, $r$, of two cities is calculated via a great circle distance algorithm. The corresponding probability density function (pdf) is shown in Fig.\ref{fig:cities}\,b). For  convenience, we used the logarithm of $r$,  $x=\log_{10}(r)$. A bin width $0.1$ in the logarithm scale was adopted to estimate the pdf. It is interesting to note that the measured pdf agrees  well with the 
Bramwell-Holdsworth-Pinton (BPH) formula \citep{Bramwell1998Nature}, which is:  
\begin{equation}
\Pi(y)=K(e^{y-e^y})^a,\,y=b(x-s),a=\pi/2,
\end{equation}
where parameters $b=0.938$, and $K=2.14$ were obtained numerically \citep{Bramwell2000PRL}. 
Note that this formula \red{was} first introduced to characterize rare fluctuations in turbulence and critical phenomena. The neighbor distance was often chosen based on cities  located near a water source.   Therefore, this neighbor  distance  could be used as a proxy of the spatial distribution of water sources. However, this postulate  needs to be verified by carefully analyzing  neighbor distance statistics for different regions. 
 Fig. \ref{fig:Example} shows the recorded PM$_{2.5}$ index with unit $\si{\mu g /m^3}$ for three typical cities,  Beijing, Shanghai and Xiamen. Visually, the measured index shows similar evolution trends: they are higher during the winter and smaller during the summer, showing an annual cycle.  In the following analysis, these database are analyzed  by  pairing two cities, i.e., $[\theta_i(t), \theta_j(t)]$ with the neighbor distance $r_{ij}$, where $\theta_i(t)$ is the air quality index of the $i$th city.

\section{Results}\label{sec:result}

\begin{figure}[!h]
\centering
\includegraphics[width=0.85\textwidth]{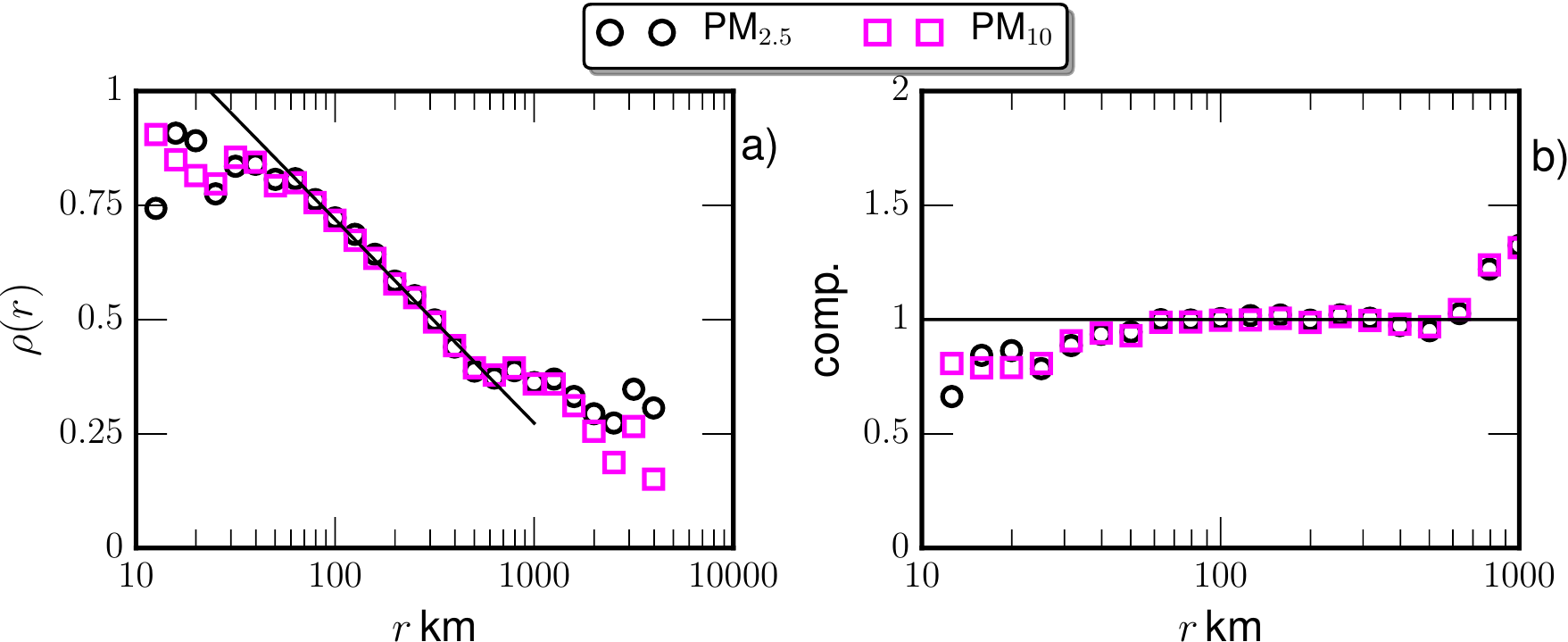}
\caption{(Color online) a) Measured spatial correlation function $\rho(r)$ of PM$_{2.5}$ ($\ocircle$) and  PM$_{10}$ ($\square$) . A log-law is observed on the range $50\le r\le 500\,$km with an scaling exponent $\beta=0.45\pm0.02$.  b) The corresponding compensated curve using the fitted parameters to emphasize the experimental log-law behavior. }
\label{fig:correlation} 
\end{figure}

\subsection{Spatial correlation}

We first calculated the spatial correlation function  for different neighbor distances $r$. The spatial correlation $\rho(r)$ is defined by the following equation:
\begin{equation}
\rho(r)=\frac{1}{N(r)}\sum^{N(r)}\frac{\left\langle \tilde{\theta}_i(t) \tilde{\theta}_j(t) \vert r_{ij}=r\right\rangle_t}{\sigma_{i}\sigma_{j}},
\end{equation}   
where $\tilde{\theta}_i(t)=\theta_i(t)-\langle  \theta_i(t) \rangle_t $ is the centered index of the $i$th city,  $\langle \,\rangle_t $ is  the time average, $\sigma_i$ is the standard deviation, $r$ is the neighbor distance; and  $N(r)$ is the number of pairs with distance $r$, where a bin width $0.1$  in the logarithm scale is used. The final $\rho(r)$ is then calculated for all pairs of cities with distance $r$.  Fig. \ref{fig:correlation}\,a) shows the measured $\rho(r)$ in a semilog plot for the PM$_{2.5}$ ($\ocircle$) and PM$_{10}$ ($\square$).  A log-law is observed in the range $50\le r\le 500\,$km, as follows:
\begin{equation}
\rho(r)\propto A -\beta\log_{10}(r),
\end{equation}
where $\beta$ is the scaling exponent, which is experimentally  $\beta=0.45\pm0.02$.
To emphasize the experimental log-law behavior, Fig. \ref{fig:correlation}\,b) shows the corresponding compensated curve using the fitted parameters.  A clear plateau confirms the existence of the log-law. Note that the log-law range is between  the microscale (resp. $1\,$km or less) and synoptic scale (resp. up to $1000\,$km),  corresponding to  the mesoscale movement in the atmospheric boundary layer \citep{Vallgren2011PRL}. 

\begin{figure}[!htb]
\centering
\includegraphics[width=0.85\textwidth]{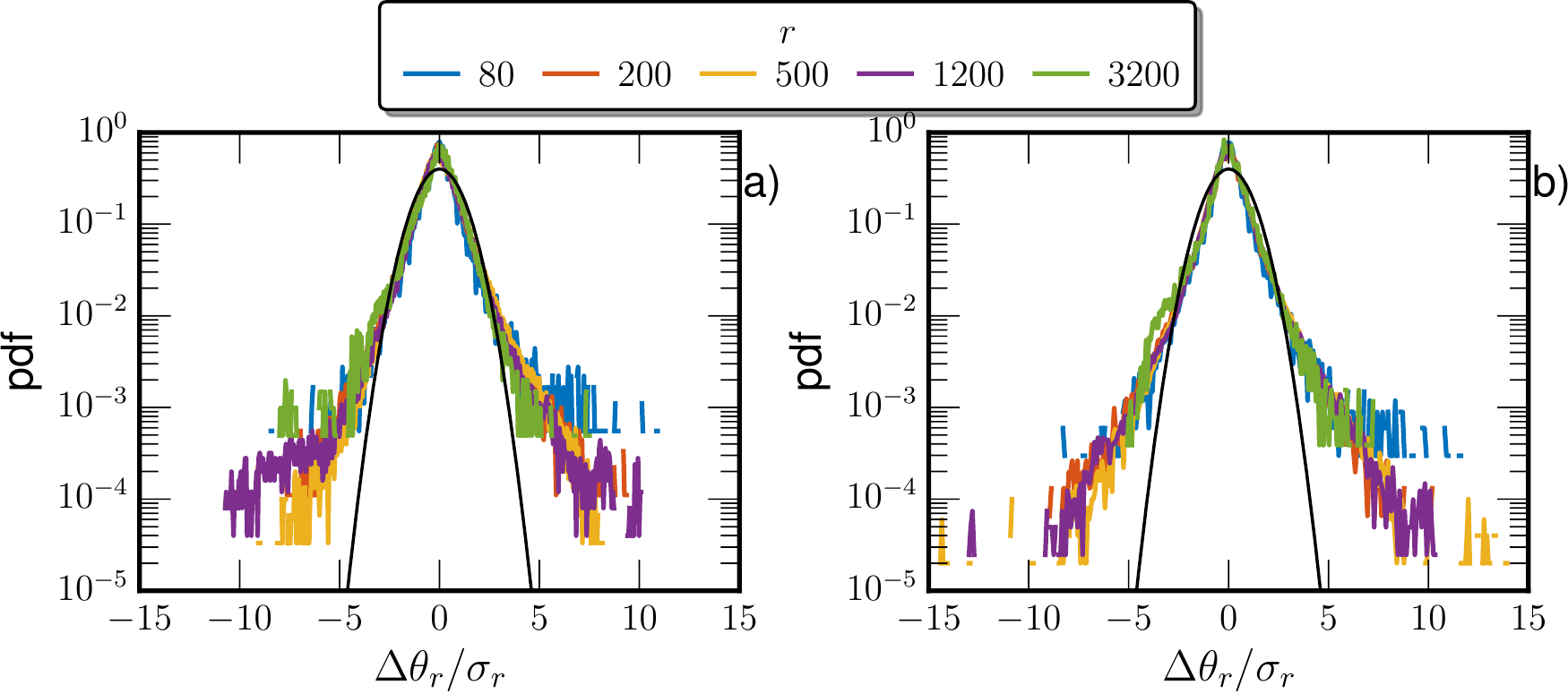}
\caption{Experimental probability density function  of the increment $\Delta \theta_r/\sigma_r$  for various separation scales $r$: a) PM$_{10}$ and b) PM$_{2.5}$.  For comparison, the normal distribution is illustrated by a thin solid line.  The core part, i.e., $-2.5\le \Delta \theta_r/\sigma_r\le 2.5$ can be fitted by an exponential law with a slope $0.67\pm0.02$. The tail, e.g., $ \vert \Delta \theta_r/\sigma_r\vert\ge 4$ has an exponential trend with  fitted slope $0.35\pm0.03$.}\label{fig:increment}
\end{figure}

\begin{figure}[!htb]
\centering
\includegraphics[width=0.85\textwidth]{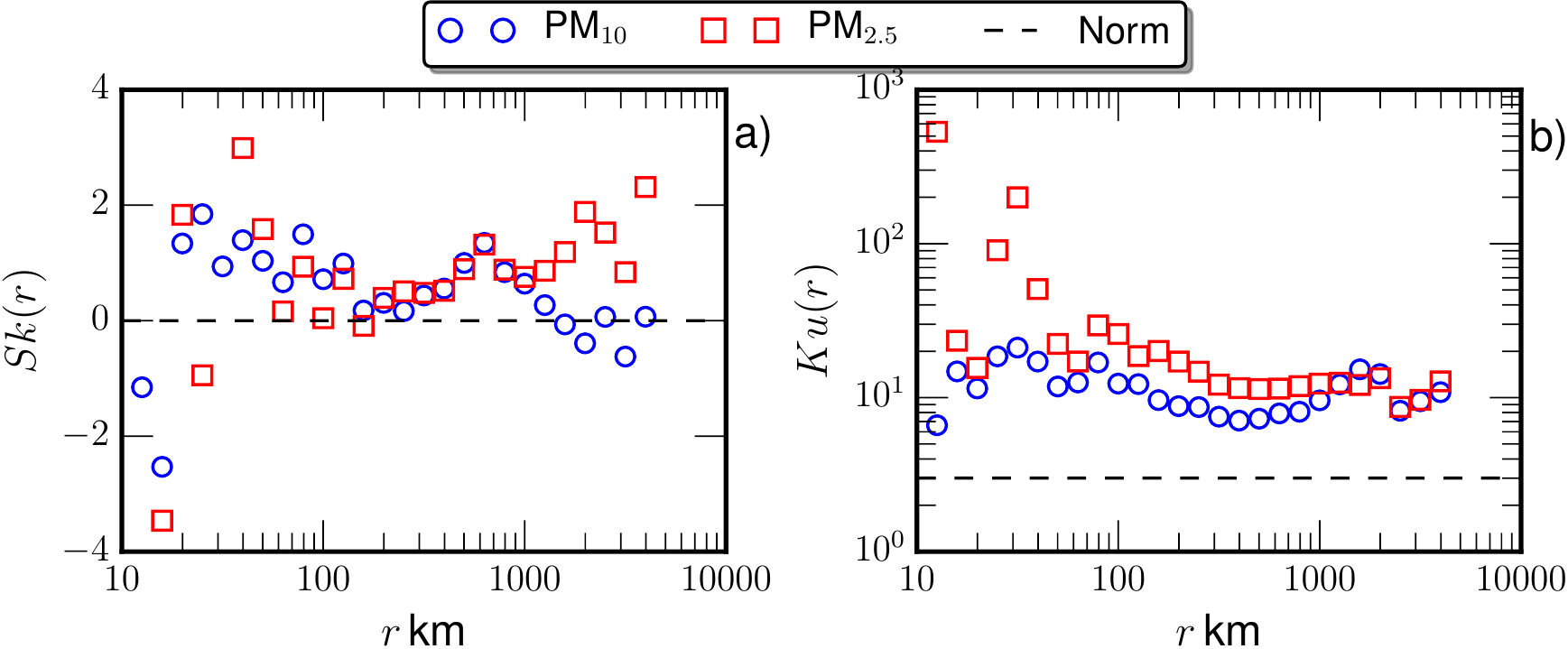}
\caption{ Measured a) skewness $Sk(r)$ and b) kurtosis $Ku(r)$ for PM$_{10}$ ($\ocircle$) and  PM$_{2.5}$ ($\square$).  For comparison, the corresponding value of the normal distribution is illustrated by a  dashed line. }\label{fig:Sk}
\end{figure}

\begin{figure}[!h]
\centering
\includegraphics[width=0.85\textwidth]{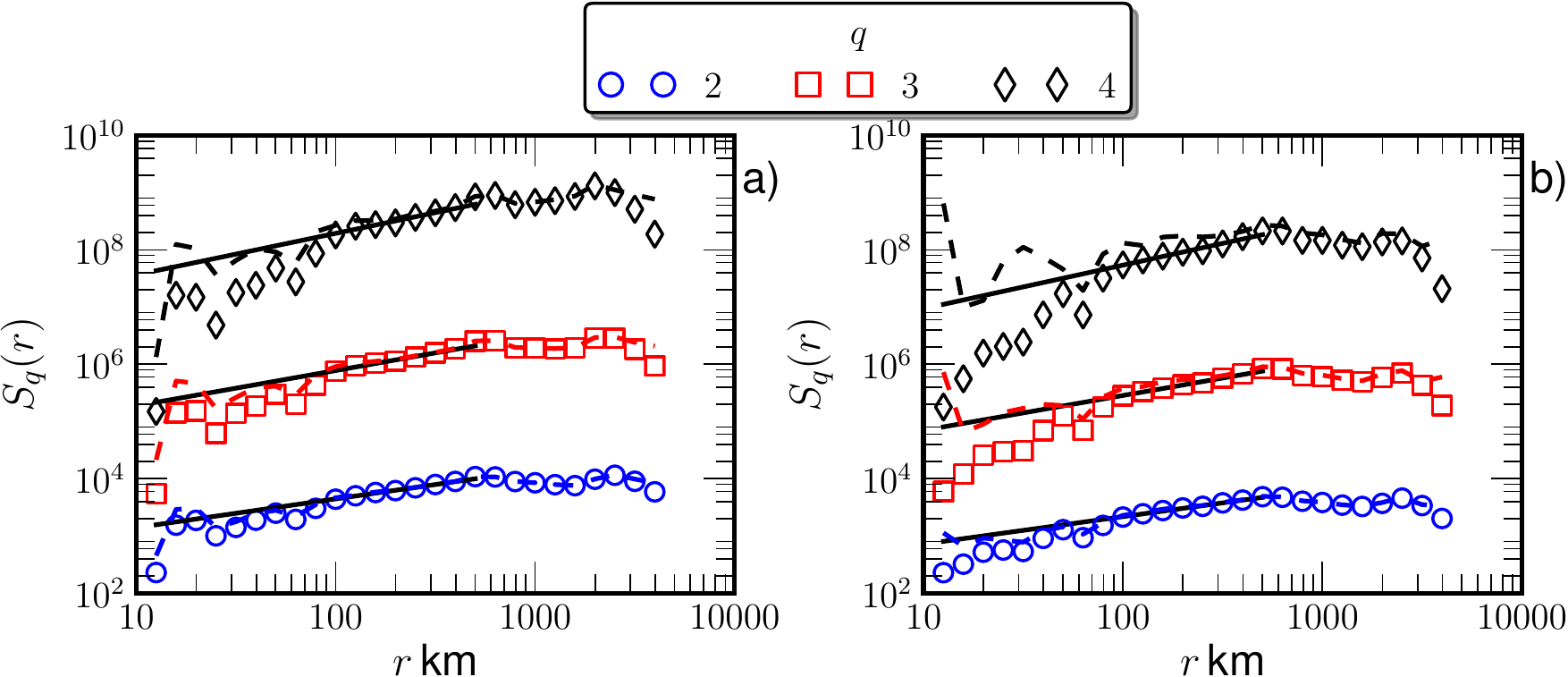}
\caption{(Color online) Measured high-order structure function $S_q(r)$: a) PM$_{10}$ and b) PM$_{2.5}$. The dashed lines are the raw structure functions. The symbols are the functions without contamination of rare events. Power-law behavior  is observed in the mesoscale  range $90\le r\le 500\,$km. The solid line is a power-law fitting. }
\label{fig:SFs} 
\end{figure}

\begin{figure}[ht]
\centering
\includegraphics[width=0.65\linewidth,clip]{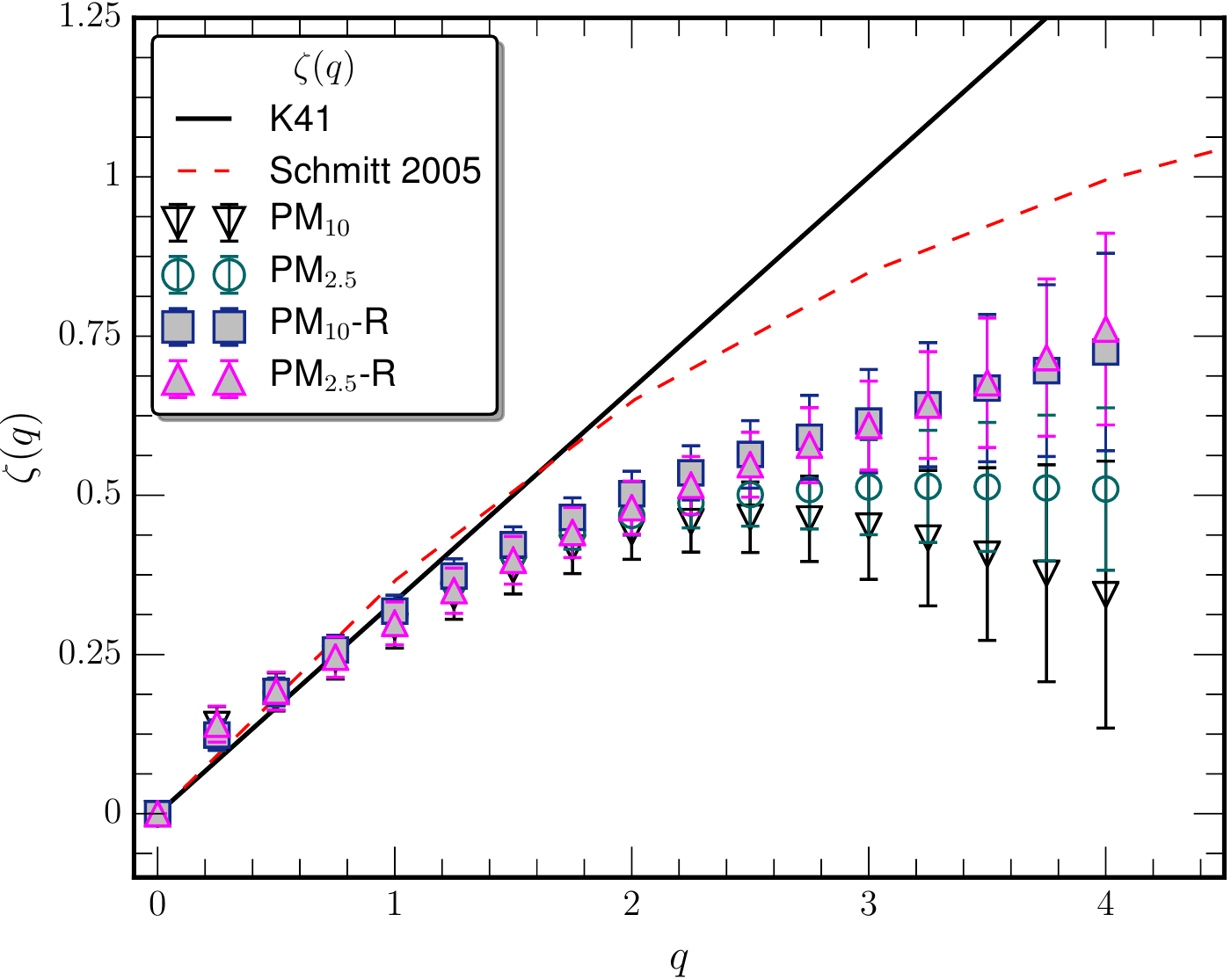}
\caption{Measured scaling exponents $\zeta(q)$ for PM$_{2.5}$ and PM$_{10}$ with and without  (denoted as PM$_{2.5}-$R or PM$_{10}-$R) rare events. For comparison, the Kolmogorov value $q/3$ and the complied  scaling exponents for the passive scalar (dashed line) are also shown.}
\label{fig:scaling}
\end{figure}

\begin{figure}[ht]
\centering
\includegraphics[width=0.65\linewidth,clip]{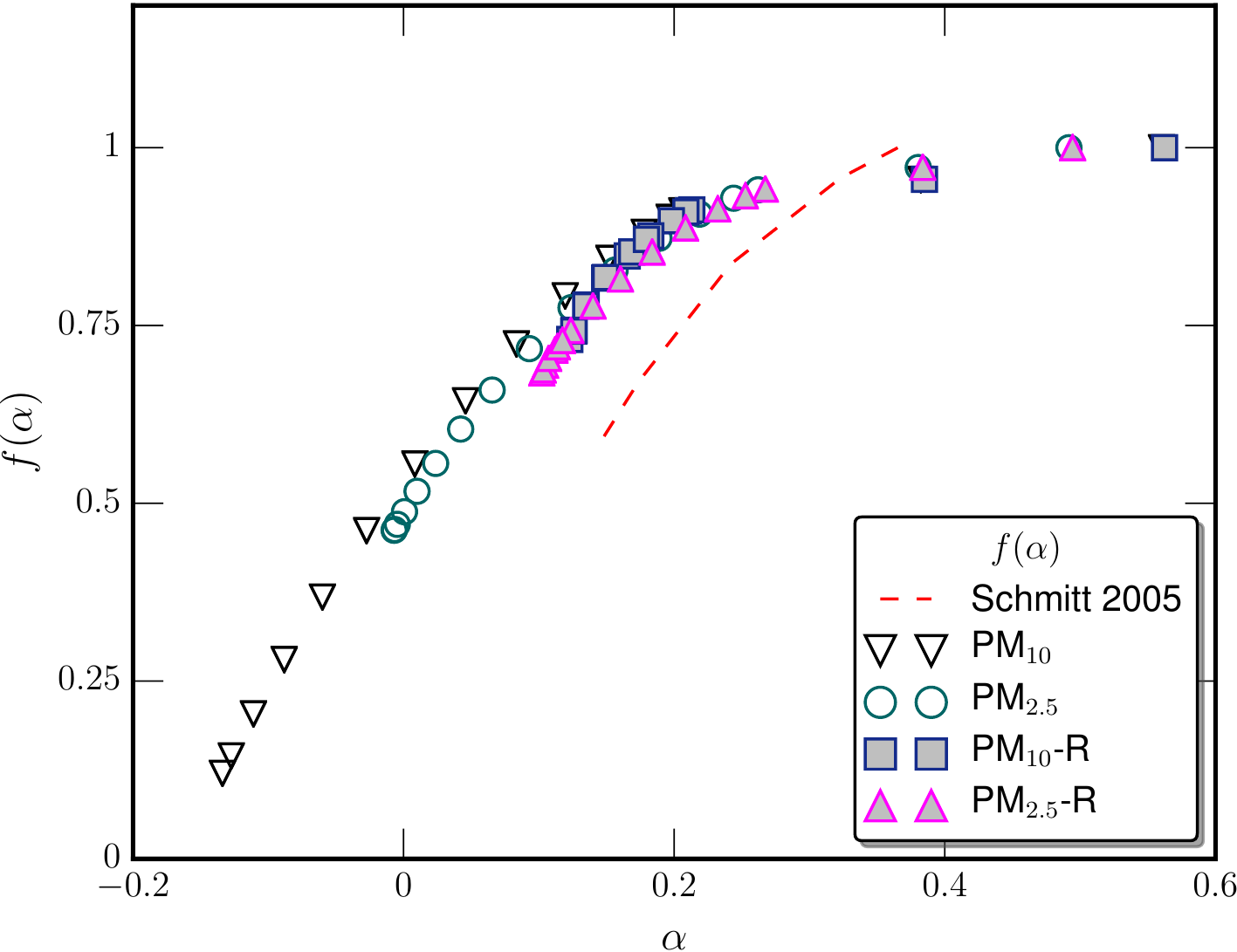}
\caption{Experimental singularity spectrum $f(\alpha)$ with and without  the rare events. For comparison, the singularity spectrum for the passive scalar on the range $0\le q \le 4$ is also presented by dashed line.}
\label{fig:singularity}
\end{figure}

\subsection{Structure Function Analysis}\label{sec:methodology}

Intermittency or multifractality is an important feature of the turbulent-like dynamical systems
\citep{Frisch1995}.  More precisely, numerous spatial or temporal freedoms exist simultaneously  and interact with each other to transfer energy, momentum, or other physical quantities.  To characterize this multiscale interaction, structure function analysis is used  to retrieve the scale invariance for high Reynolds turbulent flows \citep{Kolmogorov1941}.  It is then widely used 
in a variety of fields, including financial activity \citep{Schmitt1999,Ghashghaie1996a,Li2014PhysicaA}, crack of rock surfaces \citep{Schmittbuhl1995JGR}, rainfall patterns \citep{Tessier1996}, etc., to retrieve the scale invariant parameters. 

To characterize the interaction between different scales, the $q$th-order structure function is defined as follows:
\begin{equation}
S_{q}(r)=\left\langle  \vert \Delta\theta_{r}\vert ^q\right\rangle _{t},
\end{equation}
where $\Delta\theta_{r}=\left\{\theta_i(t)-\theta_j(t)\right\} \vert{r_{ij=r}}$ is a set of the increment; and $r$ is the separation scale (resp. neighbor distance in this work).
Note that the increment set $\Delta \theta_r$ is first calculated for all pairs with distance $r$.  Then, the $q$th-order structure function $S_q(r)$ is estimated. 
In the case of scale invariance, the structure function exhibits  a power-law behavior:
\begin{equation}
S_q(r)\sim r^{\zeta(q)},
\end{equation}
where $\zeta(q)$ is the scaling exponent.  For the turbulent velocity and passive scalar, the corresponding Kolmogorov scaling exponent without intermittent correction is $\zeta(q)=q/3$ \citep{Frisch1995}. However, both experiments and numerical simulations show that the measured $\zeta(q)$ deviates from $q/3$ \citep{Anselmet1984,Sreenivasan1997, Frisch1995,Warhaft2000,Huang2010PRE}.  It is then recognized as the intermittent nature of the dissipation field \citep{Kolmogorov1962,Frisch1995}.  A concept of multifractality is then put forward to interpret this break in the self-similarity \citep{Parisi1985,Benzi1984}. It was determined that the multifractality is one of the most important features of the turbulent-like  dynamic systems.

Fig. \ref{fig:increment} shows the measured pdf of the increment $\Delta \theta_r/\sigma_r$ on several separation scales, where $\sigma_r$ is a standard deviation of the $\Delta \theta_r$. For comparison, the normal distribution is illustrated by a thin solid line.
Roughly speaking, all these  pdfs collapse with each other, not only for different separation scales, but also for PM$_{10}$ and PM$_{2.5}$ (not shown here).   They can be further divided into three parts: the core part  ($\vert \Delta \theta_r/\sigma_r \vert\le 2.5$), the tail part with rare events  ($\vert \Delta \theta_r/\sigma_r \vert\ge4$), and  the transition range,  $2.5\le \vert \Delta \theta_r/\sigma_r \vert\le 4$. The first two parts can be fitted using an exponential distribution  with slopes $0.67\pm0.02$ and $0.35\pm0.03$, respectively. To compare to the normal distribution, we calculated the skewness and kurtosis as follows:
\begin{equation}
Sk(r)=\frac{\langle \Delta \theta_r^3\rangle }{\sigma_r^3},\,Ku(r)=\frac{\langle \Delta \theta_r^4\rangle }{\sigma_r^4}
\end{equation}
 The skewness,
$Sk(r)$, indicates an asymmetric  shape of experimental pdf, where $SK=0$ if the pdf is symmetric.  The kurtosis $Ku(r)$ is used to characterize the deviation from the normal distribution. 
 Fig. \ref{fig:Sk} shows the measured a) $Sk(r)$ and b) $Ku(r)$, where the value of the normal distribution is illustrated by a dashed line. Visually, these two statistics exhibits a scale-dependence, indicating  potential intermittency.

Fig. \ref{fig:SFs} shows the experimental high-order structure functions, $S_q(r)$,
up to $q=4$ for a) PM$_{10}$ and b) PM$_{2.5}$.
To avoid a possible contamination of rare events, the structure functions were also calculated after
removing the values where the histogram \red{contains} less then 10 data points; the corresponding $S_q(r)$ is shown as symbols.
Power-law behavior is observed in the range $90\le r\le 500\,$km. The scaling exponent $\zeta(q)$ was then calculated on this range using a least square fitting algorithm. 
Fig. \ref{fig:scaling} shows the measured scaling exponent, $\zeta(q)$, in which the errorbar indicates a $95\%$ fitting confidence.  For comparison, the Kolmogorov non-intermittent value $q/3$ (solid line) and the value for passive scalar compiled by \cite{Schmitt2005} (dashed line)
are also shown. When $q\le 2$, the measured $\zeta(q)$ are almost the same for PM$_{10}$ and PM$_{2.5}$ with or without rare events. When $q>2$, the value with rare events bends down and deviates more from $q/3$, showing  a more intermittent statistics.  We also note that due to the rare events the scaling exponents of PM$_{10}$ \red{are} more intermittent than those of PM$_{2.5}$.

\section{Discussion}\label{sec:discussion}

To characterize the intensity of the intermittency or multifractality, we calculate the singularity spectrum $f(\alpha)$ via the Legendre transform, i.e.,
\begin{equation}
\alpha=\frac{d \zeta(q)}{d q},\, f(\alpha)=\min_q\left\{ q\alpha -\zeta(q)+1 \right\},
\end{equation}
where $\alpha$ is the generalized Hurst number or intensity of multifractality. The broader  measured $\alpha$
and $f(\alpha)$ are, the more intermittent the field is \citep{Frisch1995}. This singularity spectrum was first introduced in \red{the} 1980s to characterize the multifractality of the turbulence and chaotic systems \citep{Parisi1985,Benzi1984}.  Fig. \ref{fig:singularity} shows the measured $f(\alpha)$ versus $\alpha$.  For comparison, the singularity spectrum was calculated from the compiled passive scalar scaling exponent in the range $0\le q\le 4$ and is  shown as a dashed line. Visually, the measured $\alpha$
and $f(\alpha)$ conform to the existence of the multifractality. Furthermore, by removing the rare events,  the  lower part of the singularity spectrum is suppressed since \red{it corresponds to the} scaling exponents $\zeta(q)$ shown in Fig.\,\ref{fig:scaling}. 
However, the singularity spectrum either with or without the rare events (resp. strong local events) is more intermittent than the passive scalar \citep{Warhaft2000}.

To simplify the discussion, we considered here the particulate matter as a continuous media, whereas $\theta(x,t)$ is a function of spatial coordinate, $x$, and time, $t$,
and 
an analogy to the passive scalar turbulence. The governing equation for the atmospheric particulate matter  is then  written as: 
\begin{equation}
\frac{\partial \theta(x,t)}{\partial t}+\mathbf{u}(x,t) \cdot\nabla \theta(x,t)=\kappa \nabla^2
\theta(x,t) +\mathbf{f}_{\theta}(x,t), \label{eq:GE}
\end{equation}  
where $\kappa$ is the mass diffusion coefficient; $\mathbf{f}_{\theta}(x,t)$ is the external forcing term (resp. the local sources);  $\nabla^2$ is the Laplace operator. This equation is a mass transport equation \red{indicating}  that $\theta(x,t)$ is mainly  advected by $\mathbf{u}(x,t)$.  Note that $\mathbf{f}_{\theta}(x,t)$ is different for different regions and time. For example, during  winter, one main source of the atmospheric particulate  matter is the coal consumption for  heating in the northern China. Due to the advection of the atmospheric mesoscale structures, the second-term in the l.h.s thus induces  scaling behavior. The measured  scaling exponent $\zeta(q)$ shows  more intermittent nature than the exponent for  passive scalar turbulence.  This variance could be an effect of the local emission towards   air pollution, or it could be other mechanisms that are ignored in the governing Equation\,(\ref{eq:GE}).

\section{Conclusion}\label{sec:conclusion}
In summary,  we present an analysis of the spatial statistics of particulate matter  in China in this paper.  We determined that the neighbor distance, $r$, can be described   by the BHP formula without tuning any parameters. The experimental spatial correlation function
$\rho(r)$ obeys a log-law behavior in the mesoscale range $50\le r\le 500\,\si{km}$ with an experimental scaling exponent $0.45$. 
 Additionally, the high-order structure functions obey power-law behavior  in the range   $90\le r \le 500\,\si{km}$. The retrieved scaling exponent $\zeta(q)$ curves were determined to be convex, showing \red{the} multifractal nature of the particulate matter.  Moreover, the findings from these exponents demonstrate  \red{a} more intermittent dynamics  than the exponents for passive scalar turbulence. The results presented in this paper provide a better understanding of the multiscale dynamics of air pollution, especially for the particulate matter advected by the mesoscale structures in the atmosphere.

\noindent\section*{Acknowledgements}
 This work is partially sponsored by the National Natural Science Foundation of China (NSFC) under Grant No.  11202122, 11222222, 11301433 and 11332006,  and the Fundamental Research Funds for the Central Universities (Grant No. 20720150075). {This work is partially financially supported by the Jinhua EPB's Funds (Grant No: YG2014-FW673-ZFCG046).}  The analyzed data and a  \textsc{Matlab} source code  package are available at: {https://github.com/lanlankai}.
\section*{References}
\bibliographystyle{elsarticle-harv}

\end{document}